\documentclass[letterpaper,showpacs,floatfix, aps,prb,amsmath,twocolumn,eqsecnum]{revtex4}

\usepackage{epsfig}
\usepackage{verbatim}
\usepackage{epsf}
\usepackage{array}

\newcommand{\be}{\begin{equation}}
\newcommand{\ee}{\end{equation}}
\newcommand{\nn}{\boldsymbol{n}}
\newcommand{\dnnn}{\delta\left(\widehat{\nn_{1}\nn_{2}}\right)}
\newcommand{\w}{\omega}
\newcommand{\W}{\Omega}

\renewcommand{\v}{\boldsymbol{v}}
\newcommand{\s}{{\mathrm s}}
\newcommand{\q}{\boldsymbol{q}}
\newcommand{\sgn}{\mbox{sgn}}

\renewcommand{\Re}{\mathrm{Re}}
\renewcommand{\Im}{\mathrm{Im}}

\newcommand{\req}[1]{Eq.~(\ref{#1})}
\newcommand{\reqs}[1]{Eqs.~(\ref{#1})}
\newcommand{\rref}[1]{(\ref{#1})}

\renewcommand{\L}{{\cal L}}

\newcommand{\sd}{\sigma_{\scriptscriptstyle D}}
\newcommand{\kwf}{\kappa_{\scriptscriptstyle W\!F}}
\newcommand{\cV}{{\scriptstyle C}_{\scriptscriptstyle V}}

\newcommand{\ocite}[1]{Ref.~\onlinecite{#1}}
\newcommand{\dl}{\delta\mathrm{L}}

\newcommand{\g}{{\mathrm g}}

\newcommand{\footnoteremember}[2]{\footnote{#2}\newcounter{#1}\setcounter{#1}{\value{footnote}}}
\newcommand{\footnoterecall}[1]{\footnotemark[\value{#1}]}

\begin{document}

\title{Interaction corrections: temperature and parallel field dependencies of
the Lorentz number in two-dimensional disordered metals.}

\author{G. Catelani}
\affiliation{Laboratory of Atomic and Solid State Physics, Cornell University,
Ithaca, NY 14853}

\begin{abstract}
The electron-electron interaction corrections to the transport
coefficients are calculated for a two-dimensional disordered metal in a parallel magnetic field
via the quantum kinetic equation approach. For the thermal transport, three regimes (diffusive,
quasiballistic and truly ballistic) can be identified as the temperature increases.
For the diffusive and quasiballistic regimes, the Lorentz number
dependence on the temperature and on the magnetic field is studied. The
electron-electron interactions induce deviations from the Wiedemann-Franz law, whose
sign depend on the temperature: at low temperatures the long-range part of the Coulomb interaction gives
a positive correction, while at higher temperature the inelastic collisions dominate the negative
correction. By applying a parallel field, the Lorentz number becomes a non-monotonic function
of field and temperature for all values of the Fermi-liquid interaction parameter in the diffusive
regime, while in the quasiballistic case this is true only sufficiently far from the Stoner instability.

\end{abstract}
\date{\today}
\maketitle

\section{Introduction}
\label{intro}

A standard result of the Drude-like theory of transport in disordered metals
is the Wiedemann-Franz law\cite{WF} relating the (Drude) thermal ($\kappa_{\scriptscriptstyle D}$)
and electrical ($\sigma_{\scriptscriptstyle D}$) conductivities via the Lorentz number
$\mathrm{L}_0$:
\be\label{wfl}
\frac{\kappa_{\scriptscriptstyle D}}{\sigma_{\scriptscriptstyle D} T} =
\mathrm{L}_0 \equiv \frac{\pi^2}{3e^2} \, ,
\ee
where $T$ is the temperature in energy units ($k_B=1$) and $e$ is the electronic charge.
``Drude-like theory'' means that two assumptions are made in order to calculate the transport coefficients:
1. the electrons do not interact with each other; 2. the scattering of the electrons onto the impurities
is elastic.\cite{CT,BE} While it was shown long ago\cite{AA79} that the
interplay of electron-electron interactions and disorder leads to
logarithmically divergent, temperature-dependent corrections to the electrical
conductivity at low temperatures $T\ll \hbar/\tau$ ($\tau$ is the mean free
time for the impurity scattering), it is only recently that such effects have
been correctly evaluated at higher temperatures\cite{ZNA} and for the thermal
transport.\cite{CA,Raimondi,Smith} In particular early calculations\cite{Castellani,Livanov} of the
interaction corrections to the thermal conductivity arrived at contradictory
results, due to technical difficulties in the proper construction of the energy
current density operator (both in the diagrammatic technique and in the
kinetic equation approach). This issue has been resolved in \ocite{CA},
where the local form of the collision integral for the kinetic
equation is also presented.

In deriving the quantum kinetic equation, a proper description of the disordered Fermi-liquid is obtained by
introducing bosonic soft modes (interacting electron-hole pairs) which contribute to the energy transport but,
being neutral, not to the charge transport. For interaction in the triplet channel these
bosons have a total spin $L=1$; this spin degree of freedom is affected by the magnetic field: the
description of such effects is a central part of the present work.
By extending the results of \ocite{CA}, I analyze in
detail, for two-dimensional systems, the temperature and parallel magnetic field $H$
dependencies of the ``generalized'' Lorentz number $\mathrm{L}$, defined as
\be\label{gln}
\mathrm{L}(T,H) \equiv \kappa(T,H)/\sigma(T,H) T \, ,
\ee
where, due to the electron-electron interaction corrections, both
conductivities are temperature- and field-dependent (a similar analysis for zero-dimensional
systems -- open quantum dots -- is presented in \ocite{yca}; the parallel field dependence of $\sigma$
is considered in \ocite{znapar}). Because of difficulties
in accurately measuring the electronic thermal conductivity, very few
experiments have been performed in two-dimensional systems with regards to the thermal
transport -- for example the Wiedemann-Franz law was found to hold\cite{exp1}
in a Si MOSFET sample within the experimental
accuracy, and the validity of this law was checked for the weak localization
correction.\cite{exp2} One of the difficulties in determining
$\kappa$ is in separating  the electronic contribution to
the total thermal conductivity from the phonons' contribution; however,
by measuring the thermal conductivity in the presence of a magnetic field it may be possible
to extract the electronic field-dependent part, as done
e.g. for cuprate superconductors.\cite{expcup}
Since new methods for measuring the thermal conductivity in thin films are
being explored,\cite{expnew} the study of the field dependence of $\mathrm{L}$
could be experimentally relevant.

The paper is organized as follows: in the next section I examine the
temperature dependence of the Lorentz number $\mathrm{L}$ to identify
different regimes as a function of the dimensionless parameter
$T\tau/\hbar$ and to discuss the various approximations involved.
In Sec.~\ref{pfd} I present the results for the
dependence of $\mathrm{L}$ on the parallel magnetic field.
The derivation of these results is given in Sec.~\ref{der}.
After the conclusions, I briefly consider in Appendix~\ref{appsh} the
field-dependent correction to the specific heat.
Appendix~\ref{b1der} contains some mathematical details, and Appendix~\ref{app}
a discussion of the electrical magneto-conductivity in parallel field.

\section{Temperature dependence of the Lorentz number.}
\label{td}

This section contains the result for the temperature dependence of the
Lorentz number $\mathrm{L}$.
It is convenient to separate $\mathrm{L}$ into a ``Wiedemann-Franz law'' part
$\mathrm{L}_0$ [Eq.~\rref{wfl}] and a ``violation'' part $\dl$ as
follows:
\be\label{devdef}
\mathrm{L}(T) = \mathrm{L}_0 + \delta\mathrm{L}(T) \, ,
\ee
and for the correction $\dl$ to distinguish the contributions due to interactions
in the singlet and in the triplet channels:
\be\label{dlspt}
\delta\mathrm{L} = \delta\mathrm{L}^s + \delta\mathrm{L}^t \, .
\ee
For clarity, the two terms are considered separately. The results given below are derived in
Sec.~\ref{dtd} within the quantum kinetic equation approach -- this is a
perturbative approach with $1/\g$ as the small parameter, where $\g = \sigma/(2e^2/h) \gg 1$
is the dimensionless conductance; it assumes the validity of the Fermi-liquid picture, which in turns
requires  $T\ll E_F$, with $E_F$ the Fermi energy.

\subsection{Singlet channel.}

The singlet correction $\dl^s$ is, with logarithmic accuracy:
\be\label{dls}\begin{split}
\frac{\dl^s}{{\mathrm L}_0} = &\frac{1}{\pi \g} \bigg[
 g_1\big(2\pi T \tau/\hbar\big) \ln \Big(r_s E_F/T\Big) \\
&-\frac{1}{4} g_2\big(\pi T\tau/\hbar\big) \ln \Big( 1+(\hbar/2\pi T\tau)^2\Big) \\
&-\frac{1}{5} \bigg( \frac{2\pi T\tau}{\hbar}\bigg)^2 \ln \Big( E_F/T\Big)\bigg] \, ,
\end{split}\ee
where
the functions $g_1$ and $g_2$, given in \reqs{gsdef}, describe
the crossover form the low temperature diffusive regime to the higher temperature
quasiballistic one, and $r_s$ is the ``gas parameter'' characterizing the interaction
strength:
\be
r_s= \frac{\sqrt{2}e^2}{\varepsilon\hbar v_F }
\ee
with $v_F$ the Fermi velocity
and $\varepsilon$ the dielectric constant.

In the diffusive regime $T\ll \hbar/2\pi\tau$, both $g_1$ and $g_2$ tend to 1
and \req{dls} reduces to:
\be\label{dlsd}
\frac{\dl^s_d}{{\mathrm L}_0} = \frac{1}{2\pi \g} \ln \left( \frac{\pi\hbar Dk^2}{2T}\right) \, ,
\ee
where $D=\tau v_F^2/2$ is the diffusion constant and
$k=2\pi \nu e^2/\varepsilon$ is the 2D inverse screening radius (where
$\nu=m/\pi$ is the 2D density of states).
Compared to the Altshuler-Aronov interaction correction to the electrical
conductivity,\cite{AA79} this logarithmic correction to the Lorentz number has a
completely different physical origin: it arises from the energy transported over
long distances by the neutral bosonic soft modes of the interacting electron system.\cite{CA}
At low temperatures, this additional channel for the energy transport (as compared to the
charge transport) leads to an increase in the thermal conductivity over the electrical conductivity
and therefore to an enhancement of the ``generalized'' Lorentz number (\ref{gln}).

\begin{figure}[!t]
\begin{flushleft}
\includegraphics[width=0.47\textwidth]{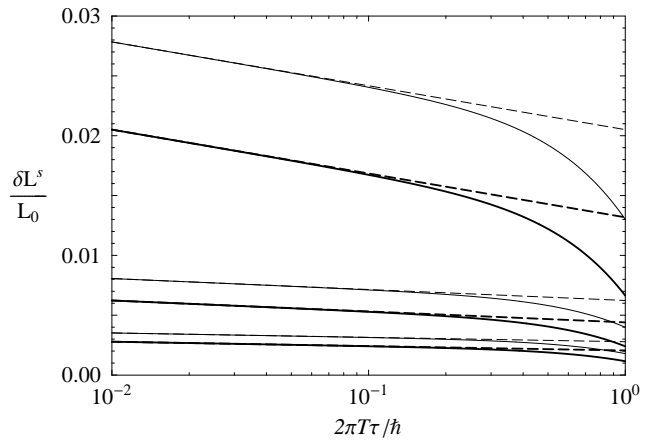}
\end{flushleft}
\caption{Relative correction to the Lorentz number as a function of temperature
in the diffusive regime. Solid lines are given by \req{dls}, while dashed lines by the
approximate expression \rref{dlsd}. For thick lines $r_s=0.1$ and for thin ones $r_s=1$; the
conductances are for each pair of curves (top to bottom): $g=100$, $g=400$ and $g=1000$.}
\label{fig1}
\end{figure}

In Fig.~\ref{fig1} I plot the relative change of the Lorentz number $\dl^s/\mathrm{L}_0$, \req{dls},
as a function of $2\pi T\tau/\hbar$ for three conductances ($g=100$, $400$ and $1000$) and for two
values of the interaction strength ($r_s=0.1$ and $r_s=1$); for comparison the curves
obtained using the approximate expression \rref{dlsd} are also shown. From the figure and the
dependence on $r_s$ in \req{dls} it follows that the deviation from the Wiedemann-Franz law
grows with the interaction strength.\footnote{The applicability of \reqs{dls} and \rref{dlsd}
is limited to values of the gas parameter far from the Wigner crystal instability, i.e.
$r_s\lesssim 37$ -- see also Sec. IIIB of \ocite{ZNA}.}
For low conductances and temperatures the (positive) change
in the Lorentz number is of the order of a few percent; unfortunately the uncertainty in
measurements of the thermal conductivity in metallic films\cite{expnew} is also of this magnitude,
making a comparison with the present theory pointless.

As the temperature increases, the inelastic collisions between the electrons and the bosons tend to inhibit
the energy transport more efficiently. The quasiballistic regime is reached in the temperature range
\be\label{qbreg}
\hbar/2\pi\tau \ll T \ll T^s_{qb} \, ,
\ee
where $T^s_{qb}$ is the solution of:
\be\label{tsqbdef}
\frac{4}{5} \pi \g \left(\frac{T^s_{qb}}{E_F}\right)^2 \ln\frac{E_F}{T^s_{qb}}=1 \, ;
\ee
for large conductances this gives:
\be
T^s_{qb} \approx E_F \sqrt{\frac{5}{2\pi\g \ln (2\pi\g)}} \, .
\ee
In this regime, the dominant contribution to the singlet correction (\ref{dls}) is:
\be\label{dlsqb}
\frac{\dl^s_{qb}}{{\mathrm L}_0} = - \frac{1}{5\pi\g} \left(\frac{2\pi T\tau}{\hbar}\right)^2
\ln \left(\frac{E_F}{T}\right) \, .
\ee
According to condition \rref{qbreg}, this expression is applicable if $T^s_{qb}\gg\hbar/2\pi\tau$; this can
be satisfied only for large enough conductances. For example at $g=14$ I find (numerically)
$T^s_{qb}\approx 0.11 E_F \approx 10 \hbar/2\pi\tau$, and
at $g=720$, $T^s_{qb}\approx 0.01 E_F \approx 50\hbar/2\pi\tau$;
in the latter case, and for larger
conductances, \req{dlsqb} can be expected to have a sufficiently large range of validity, while in the former
(and generally for small conductances) there is no quasiballistic regime.
At temperatures of the order of $T^s_{qb}$ the energy transport becomes truly ballistic in nature,
as the dominant processes responsible for the relaxation of the energy current are the
inelastic electron-boson collisions and not the electron-impurity collisions. Although
the high temperature regime $T\gtrsim T^s_{qb}$ is not considered here, it can be treated within the kinetic
equation approach.\cite{mish}

\begin{figure}[!t]
\begin{flushleft}
\includegraphics[width=0.47\textwidth]{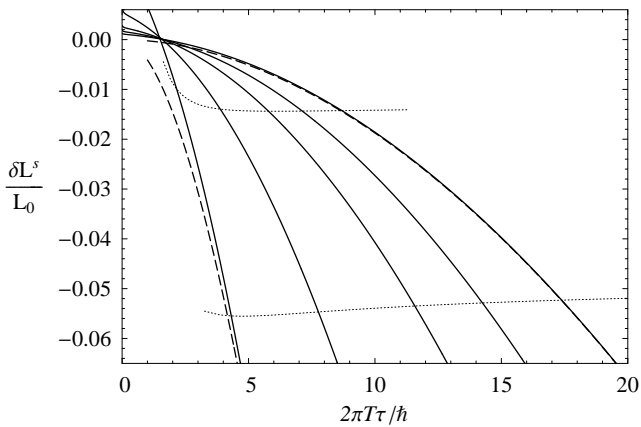}
\end{flushleft}
\caption{Relative correction to the Lorentz number as a function of temperature
in the quasiballistic regime. Solid lines are given by \req{dls} with $r_s=0.1$, while dashed lines by the
approximate expression \rref{dlsqb}. The conductances are (left to right): $g=100$, $400$, $1000$,
$1600$ and $2500$. Thin dotted lines are calculated so that they intersect the solid lines
at $T=0.1 T^s_{qb}$ (top curve) and $T=0.2T^s_{qb}$ (bottom curve).}
\label{fig2}
\end{figure}

Fig.~\ref{fig2} shows the relative change of the Lorentz number $\dl^s/\mathrm{L}_0$ as a function
of $2\pi T\tau/\hbar$ for five different conductances (solid lines); for the lowest and highest
conductances considered, the approximate result \rref{dlsqb} is also plotted (dashed lines). In agreement
with the above discussion, comparison of the dashed and solid lines shows that
\req{dlsqb} deviates significantly from the full expression \rref{dls} at low conductance, while
there is good agreement at high conductance. The
intersections between the solid lines and the upper thin dotted curve are at $T=0.1T^s_{qb}$; it is
evident that the region of validity of the quasiballistic approximation grows with the conductance.
For all conductances the (negative) correction can be a few percent; in \ocite{exp1} the Lorentz number
was measured\footnote{The experiment of \ocite{exp1} was presumably performed in the quasiballistic regime,
since the conductivity showed a linear in temperature behavior (cf. \ocite{ZNA}).} in a 2DEG and found to be slightly
smaller than ${\mathrm L}_0$, in qualitative agreement with the predictions in the present work.
However the uncertainties are
of the same order of the calculated effect and hence no quantitative comparison is possible.

\subsection{Triplet channel.}
\label{tdt}

For the triplet channel interaction correction, I consider separately, for simplicity,
the diffusive and quasiballistic regimes; in the former case the correction is:
\be\label{dltd}
\frac{\dl^t_d}{{\mathrm L}_0} = \frac{3}{2\pi\g} \ln\left(1+F_0^\sigma\right)
-\frac{1}{\pi\g} \left[1-\frac{1}{F_0^\sigma}\ln\left(1+F_0^\sigma\right)\right] \, ,
\ee
where the first term on the right hand side is again due to the bosonic energy
transport, and the second one originates from the interaction-induced
energy dependence of the elastic cross section.
While the sign of this temperature-independent correction is determined by the
sign of the Landau Fermi-liquid constant $F_0^\sigma$, its contribution
to the total correction $\dl$ [\req{dlspt}] is generally
small\footnoteremember{foot}{The correction $\rref{dltd}$ becomes large near the Stoner
instability $F^\sigma_0 = -1$; in this regime, however, approximating the Landau Fermi-liquid
parameter $F^\sigma$ by its (momentum independent) average over the Fermi surface $F^\sigma_0$
severely restricts the temperature range over which the present results can be applied
to $T/E_F\ll(1+F_0^\sigma)^2$, see \ocite{ZNA}. Therefore I do not consider this regime.} and
the overall positive sign of $\dl$ at low enough temperatures is
determined by $\dl^s_d$. I do not plot separately the contribution \rref{dltd} to $\dl$, since its
effect is simply to shift upwards (downwards) the curves in Fig.~\ref{fig1} for $F_0^\sigma>0$
($F_0^\sigma<0$).

In the quasiballistic regime the correction reads:
\be\label{dltqb}\begin{split}
\dl^t_{qb} & = 3\dl^s_{qb} \left(\frac{F_0^\sigma}{1+F_0^\sigma}\right)^2 \\
& = -\frac{3}{5\pi\g}\left(\frac{2\pi T\tau}{\hbar}\right)^2
\ln \left(\frac{E_F}{T}\right)\left(\frac{F_0^\sigma}{1+F_0^\sigma}\right)^2
\end{split}\ee
with $\dl^s_{qb}$ given in \req{dlsqb}. As for the singlet channel, this negative correction
originates from the inelastic electron-boson collision, and similarly to the singlet channel
correction, the validity of this expression is limited at high temperatures by
$T^t_{qb}$ -- the equation defining
this quantity is obtained by multiplying the left hand side of \req{tsqbdef} by
$3(F_0^\sigma/1+F_0^\sigma)^2$. It is evident that, when the quasiballistic approximation
is applicable, plotting the sum of \req{dltqb} and \rref{dlsqb} would give Fig.~\ref{fig2}
with a rescaled vertical axis, as for any value of $F_0^\sigma$ the triplet channel
contribution enhances the singlet channel correction.

\section{Parallel field dependence of the Lorentz number.}
\label{pfd}

In this section I present the results for the parallel magnetic field dependence of the
Lorentz number. The parallel field $H$ affects the electrons by shifting the energy levels
by the Zeeman energy
\be\label{ez}
E_Z = g_L \mu_B H \, ,
\ee
where $g_L$ is the Lande g-factor and $\mu_B$ the Bohr magneton.
The Lorentz number depends on $H$ only through this energy and the renormalized Zeeman energy
$E_Z^*$:
\be\label{ezs}
E_Z^* = \frac{E_Z}{1+F^\sigma_0} \,.
\ee

As it is the case for other transport properties (e.g. the magneto-conductivity),
it is convenient to consider the deviation
$\Delta {\mathrm L}$ of the Lorentz number from its zero-field value:
\be\label{Dlmdef}
\Delta {\mathrm L}(T,H) = {\mathrm L}(T,H) - {\mathrm L}(T,0) \, .
\ee
Once again I address separately the diffusive and quasiballistic regimes; in both regimes
the system is assumed to be far from the full polarization, i.e. $E_Z \ll E_F$. The derivation
of the results can be found in Sec.~\ref{dpfd}.

\subsection{Diffusive regime.}

For $T\ll \hbar/2\pi\tau$, the field-induced change in the Lorentz number is:
\be\label{Deld}\begin{split}
\frac{\Delta {\mathrm L}_d}{{\mathrm L}_0} = & -\frac{1}{\pi \g} \left(\frac{1}{F_0^\sigma} +
\frac{3}{2} \right) \left[I_1\left(\frac{E_Z}{2\pi T}\right) - I_1\left(\frac{E_Z^*}{2\pi T}\right)\right] \\
& -\frac{1}{\pi \g} \frac{1}{F_0^\sigma} \frac{E_Z}{2\pi T}
\left[I_2\left(\frac{E_Z}{2\pi T}\right)-I_2\left(\frac{E_Z^*}{2\pi T}\right) \right]
\end{split}\ee
with the dimensionless functions $I_1$ and $I_2$ defined in \reqs{I1def}-\rref{I2def}. Similarly
to \req{dltd}, the terms with the numerical prefactor $3/2$ are due to the bosonic energy
transport, while the remaining ones, proportional to $1/F_0^\sigma$, originate from the energy
dependence of the elastic cross section. The structure of this expression as the difference of
terms which depend on different energy scales (i.e. $E_Z$ and $E_Z^*$) can be traced back to
the structure of the quantum kinetic equation, in which the bosonic contributions to the collision integral
always appear as differences between a soft mode part and a ``ghost'' part.\cite{CA}

In the weak field limit $E_Z, E_Z^* \ll 2\pi T$ \req{Deld} becomes:
\be\label{Deldl}
\frac{\Delta {\mathrm L}_d}{{\mathrm L}_0} \approx -\frac{1}{\pi \g}
f_d (F_0^\sigma) \left(\frac{E_Z}{2\pi T}\right)^2
\ee
with
\be
f_d (x) = \frac{x(4+3x)}{(1+ x)^2}\, .
\ee
In the opposite case $E_Z, E_Z^* \gg 2\pi T$ the approximate formula is:
\be\label{Deldh}
\frac{\Delta {\mathrm L}_d}{{\mathrm L}_0}  \approx
-\frac{1}{\pi\g}\ln\left(1+F_0^\sigma\right)+
\frac{2}{3\pi\g}\left[1-\frac{1}{F_0^\sigma}\ln\left(1+F_0^\sigma\right)\right] \, ,
\ee
or equivalently:
\be
\Delta {\mathrm L}_d \approx -\frac{2}{3} \dl^t_d
\ee
with $\dl^t_d$ given in \req{dltd}. This result can be explained as follows:\cite{znapar}
in the diffusive regime the correction is dominated by processes with small energy and momentum exchange,
and in the strong field the bosonic modes with non-zero spin projection become gapped with the gap
given by the Zeeman energy. Therefore the contributions due to these modes must drop out from the
total correction to the Lorentz number: this is why the correction \rref{Deldh} partially cancels
the one given in \req{dltd}, with the surviving contribution originating from the modes with zero total
spin projection.

\begin{figure}[!t]
\begin{flushleft}
\includegraphics[width=0.47\textwidth]{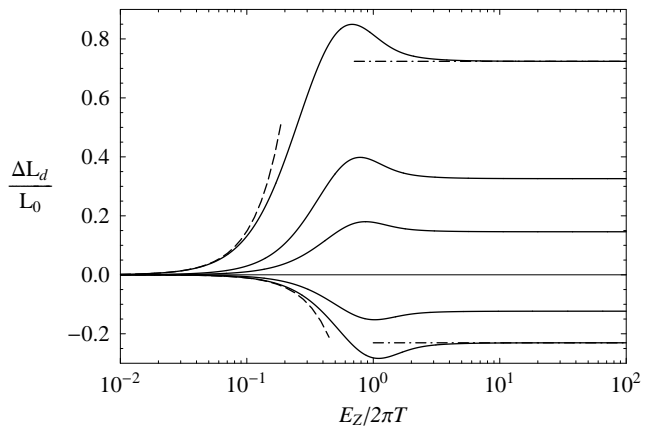}
\end{flushleft}
\caption{Relative deviation $\pi\g\Delta{\mathrm L}_d/{\mathrm L}_0$ of the Lorentz number from its zero-field
value in the diffusive regime. Solid lines are given by \req{Deld} with, from top to bottom:
$F_0^\sigma=-0.7$, $-0.4$, $-0.2$, $0.2$ and $0.4$. For $F_0^\sigma=-0.7$ and $0.4$ the approximate
expressions \rref{Deldl} (dashed lines) and \rref{Deldh} (dot-dashed) are also shown for comparison.}
\label{fig3}
\end{figure}

In Fig.~\ref{fig3} the relative deviation $\Delta {\mathrm L}_d/{\mathrm L}_0$, multiplied by $\pi g$,
is plotted as a function of $E_Z/2\pi T$ for different values of the parameter $F_0^\sigma$. At fields
such that the Zeeman energy is larger than temperature the deviation becomes quickly field-independent,
but near $E_Z\sim 2\pi T$ all the curves are non-monotonic; the presence of peaks
is due to the above discussed dependence on the two different energies $E_Z$ and $E_Z^*$, and through
the latter (and the $1/F_0^\sigma$ prefactors) the peaks' positions depend on $F_0^\sigma$.
The temperature dependence of the deviation at fixed field can also be read from
this graph by following the curves from the right (low temperature) to the left (high temperature):
the deviation is temperature independent at low temperatures $T\ll E_Z/2\pi$ and displays a
power-law decay ($\sim T^{-2}$) at high temperatures; again the non-monotonic behavior characterizes
the intermediate regime.

\subsection{Quasiballistic regime.}
\label{sec:Delqb}

Here I consider the quasiballistic regime $\hbar/2\pi\tau \ll T \ll T^t_{qb}$, with $T^t_{qb}$ defined
after \req{dltqb}. In this case I find:
\be\label{Delqb}
\frac{\Delta {\mathrm L}_{qb}}{{\mathrm L}_0} = -\frac{3}{2\pi\g}\left(\frac{2\pi T\tau}{\hbar}\right)^{\!2}
\!\left(\frac{F_0^\sigma}{1+F_0^\sigma}\right)^{\!2}
I_3\left(\frac{E_Z^*}{2\pi T};\frac{F_0^\sigma}{1+F_0^\sigma}\right)
\ee
with $I_3$ given in \req{I3def}. For $E_Z^* \ll 2\pi T$ the result takes the form:
\be\label{Delqbl}
\frac{\Delta {\mathrm L}_{qb}}{{\mathrm L}_0} \approx -\frac{1}{\pi\g} f_{qb}(F_0^\sigma)
\left(\frac{\tau E_Z}{\hbar}\right)^2 ,
\ee
where
\be\begin{split}
f_{qb}(x) & = \left(\frac{x}{1+x}\right)^2\bigg[\frac{1+2x + 4x^2}{(1+2x)^2} \\
& + \frac{2x^2(3+6x+4x^2)}{(1+2x)^3}\ln\left|\frac{x}{1+x}\right| \bigg] \, ,
\end{split}\ee
while with logarithmic accuracy the large field limit $E_Z^* \gg 2\pi T$ is:
\be\label{Delqbh}
\frac{\Delta {\mathrm L}_{qb}}{{\mathrm L}_0}  \approx \frac{2}{5\pi\g}
\left(\frac{2\pi T\tau}{\hbar}\right)^2 \ln \left(\frac{E_Z}{T}\right)
\left(\frac{F_0^\sigma}{1+F_0^\sigma}\right)^2 .
\ee
Comparison of \req{Delqbh} with \req{dltqb} shows that the partial cancelation
that was found in the diffusive limit is also
realized in the quasiballistic one, but with an important difference: the gapped modes still
contribute to the total correction to the Lorentz number because the quasiballistic corrections are
dominated by the inelastic scattering with large momentum exchange. The gap therefore excludes the
low energy ($E< E_Z$) contributions, but the higher energy ones ($E_Z<E<E_F$) are still relevant.

\begin{figure}[!t]
\begin{flushleft}
\includegraphics[width=0.47\textwidth]{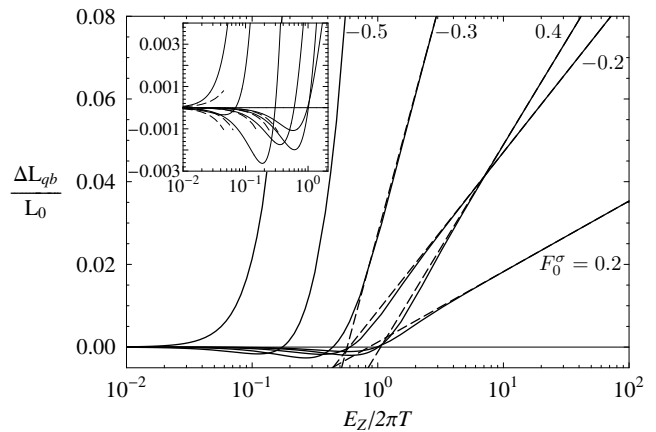}
\end{flushleft}
\caption{Relative deviation of the Lorentz number from its
zero-field value as a function of the Zeeman energy in the
quasiballistic regime (normalized as explained in the text) for
different values of the Fermi-liquid parameter, as labeled (the
curve partially covered by the inset corresponds to $F_0^\sigma
=-0.7$). Solid lines are given by \req{Delqb}, while dashed lines by
the approximate expression \rref{Delqbh}. The inset shows details of
the low-field regime, with the dashed curves corresponding to
\req{Delqbl}.} \label{fig4}
\end{figure}

\begin{figure}[!t]
\begin{flushleft}
\includegraphics[width=0.47\textwidth]{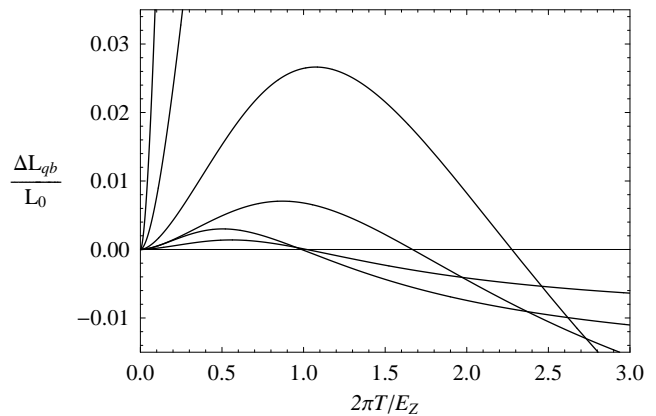}
\end{flushleft}
\caption{Relative deviation of the Lorentz number from its zero-field value as a function of the
temperature in the quasiballistic regime (normalized as explained in the text) for different
values of the Fermi-liquid parameter. Proceeding clockwise near the
origin, starting with the steepest curve, the parameters are: $F_0^\sigma=-0.7$, $-0.5$, $-0.3$,
$-0.2$, $0.4$ and $0.2$}
\label{fig5}
\end{figure}

Fig.~\ref{fig4} shows the field dependence of $\Delta {\mathrm L}_{qb}$, \req{Delqb}, by plotting
\[
\frac{\Delta {\mathrm L}_{qb}}{{\mathrm L}_0}\bigg/\frac{3}{2\pi\g}\left(\frac{2\pi T\tau}{\hbar}\right)^2
\]
as a function of $E_Z/2\pi T$; for comparison the approximate result \rref{Delqbh} is also
plotted.\footnote{This equation gives the slope of the dashed straight
lines; a subleading constant, calculated numerically, has been added to each line.} In the inset the
low-field behavior of \req{Delqb} is compared to \req{Delqbl}; except for the case $F_0^\sigma=-0.7$,
all curves are non-monotonic. The threshold value $F^\sigma_{th}$ above which
$\Delta {\mathrm L}_{qb}$ is a non-monotonic function can be found by requiring $f_{qb}(F^\sigma_{th})=0$;
this gives $F^\sigma_{th}\simeq -0.679$. Although the present results are not valid close to
the Stoner instability (see footnote\footnoterecall{foot}), they suggests that as $F_0^\sigma\to -1$ the
relationship between energy and charge transport properties can be qualitatively altered compared to
the weakly interacting case. For completeness, I consider in Fig.~\ref{fig5} the temperature dependence of
$\Delta {\mathrm L}_{qb}$ by plotting
\[
\frac{\Delta {\mathrm L}_{qb}}{{\mathrm L}_0}\bigg/\frac{3}{2\pi\g}\left(\frac{E_Z \tau}{\hbar}\right)^2
\]
as a function of $2\pi T/E_Z$ in the low to intermediate temperature regime.

\section{Derivation.}
\label{der}

This section is devoted to the calculation of the interaction
corrections to the Lorentz number using the formalism of \ocite{CA}.
In the absence of the magnetic field, one can use directly the results of
that reference, while a generalization is needed for the parallel field case,
as discussed in Sec.~\ref{dpfd}. From now on, I set $\hbar =1$.

\subsection{Temperature dependence.}
\label{dtd}

The results presented in Sec.~\ref{td} are a straightforward
consequence of the findings of \ocite{CA}, where it is shown that
for two-dimensional systems the thermal conductivity can be written as:
\be\label{kt}
\kappa= \kwf + \Delta\kappa \, .
\ee
Here
\be\label{kwf}
\kwf = \mathrm{L}_0 \sigma T
\ee
follows the Wiedemann-Franz law with $\mathrm{L}_0$ defined in \req{wfl} and the
electrical conductivity $\sigma$ includes the interaction corrections. The additional
term $\Delta\kappa = \Delta\kappa^s+3\Delta\kappa^t$ is given by the sum of the singlet
and triplet channel corrections. The former was calculated with logarithmic
accuracy:\footnote{Here an algebraic mistakes in \ocite{CA} is corrected; this correction leads to
different numerical factors which, however, do not modify qualitatively the original results.}
\be\label{dks}\begin{split}
\Delta\kappa_s = & \
\frac{T}{6} g_1 \left(2\pi T\tau \right)
\ln \left( \frac{ v_F k}{T} \right)
\\ &
- \frac{T}{24} g_2\left(\pi T\tau \right)
\ln \left( 1+ \frac{1}{(2\pi T\tau)^2} \right) \\
& -\frac{2\pi^2}{15}T
\left(T\tau \right)^2 \ln\left(\frac{E_F}{T}\right)
\end{split}
\ee
with the functions $g_1$ and  $g_2$ given by:
\begin{subequations}\label{gsdef}
\be
g_1(x) =\frac{3}{x^2}\left\{ \frac{1}{x}\Big[ 2\psi'\big(1/x\big) - x^2\Big]-2\right\} ,
\ee
where $\psi'$ is the derivative of the digamma function, and
\be
g_2(x) = \frac{26}{15} x^2 +\frac{8}{3}g_1(x) -\frac{5}{3} \, .
\ee\end{subequations}
Note that $g_1(0)=g_2(0)=1$, and $g_1(x)\simeq 3/x$ for $x\ll 1$.

Using the definitions (\ref{gln}), (\ref{kt}) and \rref{kwf} I find at first order in $1/\mathrm{g}$:
\be\label{ltl0}
\frac{\mathrm{L}(T)}{\mathrm{L}_0} = 1 +\frac{\Delta \kappa}{\sd T}
\ee
Then from the definition \rref{devdef} and \req{dks} one arrives at \req{dls}. The limiting expressions
given in \reqs{dlsd} and \rref{dlsqb}
are found by keeping the leading order contributions to $\delta{\mathrm L}^s$ for
small and large $T\tau$ respectively by using the asymptotic forms of the functions $g_1$ and $g_2$ given
after \reqs{gsdef}. The condition \rref{tsqbdef} is obtained by equating the
(absolute value of the) correction \rref{dlsqb} to the non-interacting Lorentz
number ${\mathrm L}_0$ [\req{wfl}]; the results are not valid at temperatures higher
than $T^s_{qb}$ because in solving the kinetic equation it was assumed
that the impurity scattering is the dominant process contributing to the
energy relaxation rate -- see the discussion at the end of Sec.~6.2 of \ocite{CA}; however, the
kinetic equation itself is still valid.

For the triplet channel, the correction $\Delta\kappa^t$ was calculated in the
diffusive and quasiballistic regimes:
\be\label{dktd}
\Delta\kappa^t_d =-\frac{T}{18} \left[ 1- \frac{1}{F_0^\sigma}
\ln \left( 1+F_0^\sigma \right) \right]
+ \frac{T}{12} \ln \left(1+F_0^\sigma \right)
\ee
for $T\tau \ll 1/2\pi$, and
\be\label{dktqb}
\Delta\kappa^t_{qb} = -\frac{2\pi^2}{15}T \left( T\tau \right)^2
\ln\left(\frac{E_F}{T}\right)
\left( \frac{F_0^\sigma}{1+F_0^\sigma}\right)^2
\ee
for $T\tau \gg 1/2\pi$.
Multiplying \reqs{dktd}-\rref{dktqb} by 3 and using \req{ltl0} and the definition \rref{devdef} of $\delta\mathrm{L}$,
the results \rref{dltd} and \rref{dltqb} are obtained; the factor of 3
arises from the summation over the three projections of the total spin, which in the
absence of magnetic field contribute equally to the thermal conductivity.

\subsection{Parallel field dependence.}
\label{dpfd}

To obtain the results of Sec.~\ref{pfd} I give here an extension of the calculations
of \ocite{CA}. As in \req{kt}, I separate the thermal conductivity in a part
which follows the Wiedemann-Franz law and a correction; both term now depend on the
applied parallel magnetic field $H$, or more precisely on the Zeeman splitting, \req{ez}. The term
$\kwf(T,H)$ is straightforwardly calculated using the results of \ocite{znapar}, so one needs to
consider only the correction $\Delta \kappa(T,H)$. As discussed in e.g. Refs.~\onlinecite{znapar}
and \onlinecite{AA85},
the singlet and the triplet $L_z=0$ contributions to $\kappa$ are not affected by the parallel
field [$L_z$ is the projection of the total spin along the field direction]. The effect of the field
on the remaining $L_z=\pm 1$ components of the triplet channel correction is to shift the frequency of
the interaction propagators; in the langauge of \ocite{CA}, the bosonic propagators
$\L^\sigma(\w,\q;\nn_1,\nn_2;L_z)$ are:
\be\label{ls}\begin{split}
&\L^\sigma (\nn_1,\nn_2;L_z)= \W_2 \dnnn \L_0 (\nn_1;L_z) + \\
&\ \L_0(\nn_1;L_z)\L_0(\nn_2;L_z)
\frac{\left(-i\w\frac{F_0^\sigma}{1+F_0^\sigma}+\frac{1}{\tau}\right){\cal C}(L_z)}{{\cal C}(L_z)-
\left(-i\w\frac{F_0^\sigma}{1+F_0^\sigma}+\frac{1}{\tau}\right)} \, ,
\end{split}\ee
and the corresponding (triplet) ``ghost'' propagators
$\L^g(\w,\q;\nn_1,\nn_2;L_z)$ are:\footnote{The ghosts corresponding to the singlet channel are,
of course, unchanged.}
\be\label{lg}\begin{split}
{\cal L}^g (\nn_1,\nn_2;L_z)= \W_2 \dnnn \L_0 (\nn_1;L_z) + \\
\L_0(\nn_1;L_z)\L_0(\nn_2;L_z)
\frac{\frac{1}{\tau}{\cal C}(L_z)}{{\cal C}(L_z)-\frac{1}{\tau}} \, ,
\end{split}
\ee
where
\be\label{lcdef}\begin{split}
\L_0 (\nn_i;L_z) = \frac{1}{-i(\w-L_zE_Z^*) + i \v_i\cdot\q +1/\tau}\ , \\
{\cal C} (L_z) = \sqrt{\big(-i(\w-L_zE_Z^*)+1/\tau\big)^2+\big(v_Fq\big)^2} \, .
\end{split}\ee
In the above formulas I dropped the variables $\w$, $\q$ for compactness,
$\v_i = v_F \nn_i$ and  $E_Z^*$ is
the Zeeman energy renormalized by the interactions, \req{ezs}.

The calculation of the field-dependent thermal conductivity proceeds now as in \ocite{CA}:
the evaluation of the transport coefficients can be reduced to
integrals over the frequency $\w$ whose integrands consist of a distribution function part
times a kernel part ${\mathrm K}(\w)$; the latter is found after integration over the momentum $\q$ and
summation over the total spin projections. The field-dependent
kernels can be found using the expressions \rref{ls}-\rref{lg} instead of their zero-field counterparts;
below I calculate explicitly the correction $\Delta\kappa_{m}$ to the thermal magnetoconductivity
$\kappa_{m}(T,H)=\kappa(T,H)-\kappa(T,0)$. In other words, I want to write $\kappa(T,H)$ in the form
\rref{kt} and define $\Delta\kappa_m$ as:
\be\label{Dkmdef}
\Delta\kappa_{m} (T,H) = \Delta\kappa (T,H) - \Delta\kappa (T,0) \ ,
\ee
where $\Delta\kappa(T,0)$ is the correction considered in the previous section. To calculate
$\Delta\kappa_{m}$ I introduce for each kernel ${\mathrm K}$ the corresponding kernel
difference $\Delta{\mathrm K}={\mathrm K}(H) - {\mathrm K}(0)$ between the kernel calculated
with and without the field. As in the preceding section, I consider separately the diffusive
and quasiballistic regimes.

\subsubsection{Diffusive regime.}

In the diffusive regime, the two main contributions to $\Delta\kappa$ come from the energy
dependence of the elastic cross section and from the bosonic energy transport. Writing
\be
\Delta\kappa_{m,d} = \delta\kappa_{el,m} + (\kappa^\sigma_m - \kappa^g_m) \, ,
\ee
the two terms are given by [cf. Eqs. (6.11b) and (6.36b) of \ocite{CA}]:
\begin{subequations}\label{dkdeqs}
\be
\delta\kappa_{el,m} = -2 \frac{\sd}{e^2 T} \int\!d\w \, \Delta{\cal E}(\w)
\left[\frac{\w^3}{12}\frac{\partial N_P}{\partial \w}\right]
\ee
and
\be
\kappa^\sigma_m - \kappa^g_m = \frac{\sd}{e^2 T}\int\!d\w \, \Delta{\cal B}^0(\w)
\left[\frac{\w^3}{4}\frac{\partial N_P}{\partial \w}\right]
\ee\end{subequations}
with the kernels\begin{subequations}\label{dkerd}
\be\label{Deker}\begin{split}
\Delta{\cal E} = -\frac{e^2}{\sd 2\pi^2 \w^2}&\frac{1}{F_0^\sigma}\bigg[
\omega \ln \left| \frac{\w^2-E_Z^2}{\w^2-E_Z^{*2}}\right| \\ &+ E_Z
\ln \left|\frac{\w+E_Z}{\w-E_Z}\cdot\frac{\w-E_Z^*}{\w+E_Z^*} \right| \bigg]
\end{split}\ee
and
\be
\Delta{\cal B}^0 = \frac{e^2}{\sd 2\pi^2 \w}
\ln \left| \frac{\w^2 - E_Z^2}{\w^2-E_Z^{*2}} \right| \, .
\ee\end{subequations}
As discussed above, these kernels are found by substituting the expressions
\rref{ls}-\rref{lg} for the propagators into the definitions of ${\cal E}$ and
${\cal B}^0$ given in Eqs.~(6.9) and (6.36d) of \ocite{CA}.\footnote{Alternatively, one
can use directly the diffusive limit expressions for these kernels, Eqs.~(6.22) and
(6.24) of \ocite{CA} with the appropriately modified diffusive limit propagators, which are obtained
by shifting the frequency $\w \to \w - L_zE_Z^*$.}
Substituting \reqs{dkerd} into \reqs{dkdeqs} [and by a change of variable
$\w\to 2\pi T\w$] I get:
\begin{subequations}\label{thmcd}
\be\label{dkelm}\begin{split}
\delta\kappa_{el,m} = &-\frac{T}{6}\frac{1}{F_0^\sigma} \bigg\{
I_1 \left(\frac{E_Z}{2\pi T}\right) - I_1\left(\frac{E_Z^*}{2\pi T}\right)
\\ & + \frac{E_Z}{2\pi T} \left[ I_2 \left(\frac{E_Z}{2\pi T}\right) -
I_2 \left(\frac{E_Z^*}{2\pi T}\right) \right]\! \bigg\}
\end{split}\ee
and
\be
\kappa^\sigma_m - \kappa^g_m = - \frac{T}{4} \left[
I_1\left(\frac{E_Z}{2\pi T}\right) - I_1\left(\frac{E_Z^*}{2\pi T}\right) \right] \, ,
\ee
\end{subequations}
where I introduced the dimensionless functions:
\be\label{I1def}
I_1(E) = \pi\!\int\!d\w \, \frac{\w^2}{\sinh^2 \pi\w} \ln \left| 1- \frac{E^2}{\w^2} \right|
\ee
and
\be\label{I2def}
I_2 (E) = \pi\!\int\!d\w \, \frac{\w}{\sinh^2 \pi\w} \ln \left| \frac{1+\w/E}{1-\w/E}\right| \, .
\ee
Eq.~\rref{I1def} can be identically rewritten as:
\be\label{I1a}
I_1(E) = 2 E^2 - 4\sum_{n=1}^{\infty} n\!\left[
\ln \Big(1+\big(E/n\big)^2\Big) -\frac{(E/n)^2}{1+(E/n)^2} \right]
\ee
which is useful to obtain the $E\ll 1$ expansion, and as:
\be
I_1(E) = \frac{2}{3} \ln |E| + C + \pi\!\int\!d\w\, \frac{\w^2}{\sinh^2 \pi\w}
\ln \left| 1- \frac{\w^2}{E^2}\right|
\ee
which gives the $E\gg 1$ asymptotic behavior; the constant $C$ appearing above is:
\be\begin{split}
C & \equiv -2\pi\!\int\!d\w \, \frac{\w^2}{\sinh^2 \pi\w} \ln |\w| \\
& = \frac{2}{3}\left(\gamma + \ln 2\pi -\frac{3}{2}\right) - \frac{4}{\pi^2} \zeta'(2)
\simeq 0.99 \, ,
\end{split}\ee
where $\gamma \simeq 0.577$ is the Euler constant and $\zeta'(2)\simeq -0.938$ is the derivative of the zeta
function evaluated at 2.
As for \req{I2def}, in the given form the
large $E$ limit can be readily obtained [$I_2(E)= 2/3E +\ldots$], while to find the
small $E$ limit I rewrite it as:
\be\label{I2a}\begin{split}
I_2(E) & =  \pi \, \sgn \, E -4 E \\ &+ 4 \sum_{n=1}^{\infty} \Im \left[
\ln \left(1+i \frac{E}{n}\right) -\frac{1}{1-iE/n}\right] \, .
\end{split}\ee
Note also the identity [primes indicate derivatives]:
\be
I_1'(E)+EI_2'(E) = 0
\ee
which enables to verify that the correction $\delta\kappa_{el,m}$ of \req{dkelm} vanishes
in the limit $F_0^\sigma \to 0$, as expected.

Using \reqs{thmcd}, together with the definitions \rref{kt}, \rref{Dkmdef} and \rref{Dlmdef},
and dropping terms of higher order in $1/\g$, one arrives at \req{Deld}. The approximate expressions
\rref{Deldl}-\rref{Deldh} follow from \reqs{I1a}-\rref{I2a}.

\subsubsection{Quasiballistic regime.}

In the quasiballistic regime, the correction $\Delta \kappa$ is determined by the inelastic
electron-boson collisions and it can be written as [cf. Eq. (6.39a) of \ocite{CA}]:
\be\label{dkmqb}
\Delta\kappa_{m,qb} = \frac{\sd}{e^2 T} \int\!d\w \, \Delta{\cal B}^1(\w)
\left[\frac{\w^3}{4} \frac{\partial N_P}{\partial\w}\right]
\ee
with\begin{subequations}\label{b1eqs}
\be\label{b1eq}
\Delta{\cal B}^1 = \frac{e^2}{\sd 2\pi^2} \tau^2\w \left(\frac{F_0^\sigma}{1+F^\sigma_0}\right)^2
J\left(\frac{E_Z^*}{\w};\frac{F_0^\sigma}{1+F_0^\sigma}\right) \, ,
\ee
\be\begin{split}
J(E;F) = -2\!\!\sum_{L_{z}=\pm 1}\!\bigg\{\ln \left| 1-L_z E\right| \frac{(1-L_zE)^2}{(1-L_zE)^2-F^2}
\\-  \ln \left|F\right| F^2
\left[\frac{1}{(1-L_z E)^2-F^2} - \frac{1}{1-F^2}\right] \!\bigg\} \, .
\end{split}\ee\end{subequations}
Some details on the derivation of this kernel are given in Appendix~\ref{b1der}. Substitution of
\reqs{b1eqs} into \req{dkmqb} results in
\be
\Delta\kappa_{m,qb} = - \frac{T}{4}(2\pi T\tau)^2 \left(\frac{F_0^\sigma}{1+F_0^\sigma}\right)^2
I_3\left(\frac{E_Z^*}{2\pi T};\frac{F_0^\sigma}{1+F_0^\sigma}\right)
\ee
with the dimensionless function $I_3$ defined as:
\be\label{I3def}
I_3 (E;F) = \pi\!\int\!d\w \, \frac{\w^4}{\sinh^2 \pi \w}\, J \left( \frac{E}{\w} ; F \right) \, .
\ee
The large and small $E$ limits of $I_3$ can be found by keeping the leading order terms in the expansion
of the function $J(E;F)$.\footnote{The calculation of the subleading terms is more involved for $I_3$
compared to the expansion of $I_1$ or $I_2$ and I do not attempt it here.} In this way I obtain:
\be\label{I3h}
I_3(E;F) \approx -\frac{4}{15} \ln |E|
\ee
for $E\gg 1$, and
\be\label{I3l}
I_3(E;F) \approx \frac{2}{3} E^2 \left[\frac{1+3F^2}{(1-F^2)^2}
+\frac{F^2(3+F^2)}{(1-F^2)^3} \ln F^2 \right]
\ee
for $E\ll 1$. Knowing $\Delta\kappa_{m,qb}$ and the approximate formulas \rref{I3h}-\rref{I3l},
proceeding as in the previous subsection finally leads to \req{Delqb}, \rref{Delqbl} and \rref{Delqbh}.

\section{Conclusions.}

The quantum kinetic equation approach is a powerful method to investigate the effects
of the electron-electron interactions on transport in disordered metals\cite{ZNA,CA,znapar}
and open quantum dots.\cite{yca} Using this approach I considered the temperature and
parallel magnetic field dependence of the Lorentz number in two-dimensional disordered metals.

Three regimes can be distinguished as the temperature increases: diffusive, quasiballistic and
truly ballistic. In the low-temperature diffusive regime, the Lorentz number is enhanced
above its Drude value due to the energy transported by neutral bosonic modes that describe
the interacting electron-hole pairs, see \req{dlsd} and Fig.~\ref{fig1}. At intermediate temperatures
(the quasiballistic regime) the Lorentz number is suppressed by the inelastic electron-boson collision,
\req{dlsqb} and Fig.~\ref{fig2}, with the crossover between the two regimes described by \req{dls}.
The effect of the interaction in the triplet channel is given in \reqs{dltd} and \rref{dltqb}
for the diffusive and quasiballistic regimes respectively.

If a magnetic field is applied parallel to the two-dimensional metal, the Zeeman splitting of the
electronic energies affects the transport properties; in particular in the diffusive regime the
deviation of the Lorentz number from its zero-field value displays a non-monotonic dependence
on the ratio between the Zeeman energy and the temperature, see \req{Deld} and Fig.~\ref{fig3}.
Finally, as discussed in Sec.~\ref{sec:Delqb}, in the quasiballistic regime the deviation can be either a
monotonic or non-monotonic function of both temperature and Zeeman energy depending on the value
of the Fermi-liquid parameter.

\acknowledgments

I am grateful to I. Aleiner and Y. Ahmadian for reading and commenting on the manuscript.
Financial support through the J.A. Krumhansl Postdoctoral Fellowship is acknowledged.

\appendix

\section{Parallel field dependence of the specific heat.}
\label{appsh}

In this appendix I calculate for completeness the correction $\delta\cV$ to the specific heat in the
presence of a parallel magnetic field; the general expression for $\delta\cV$ is:\cite{CA}
\be\label{sphe}
\delta\cV = \frac{\partial}{\partial T} \Big(u^\sigma - u^g \Big) \, ,
\ee
where the energy densities $u^\alpha$ are given by:
\be\label{enden}
u^\alpha = \int\!d\w \, \w \, b^\alpha(\w) N_P (\w) \, .
\ee
Here $N_P(\w)$ is the Planck distribution and $b^\alpha(\w)$ are the bosonic densities of states.
In the presence of the parallel field I find:
\be\begin{split}
&b^\sigma(\w;H) - b^g(\w;H) = \\
&-\frac{\Re}{2\pi}\!\sum_{L_{z}}\!\int\!\frac{d^2q}{(2\pi)^2} \bigg[ \frac{F_0^\sigma}{1+F_0^\sigma}
\bigg(\frac{1}{{\cal C}(L_z)-b} - \frac{1}{{\cal C}(L_z)}\bigg) \\
&- \frac{-i\left(\w - L_zE_Z^*\right)+1/\tau}{{\cal C}(L_z)}\left(\frac{1}{{\cal C}(L_z)-b} -
\frac{1}{{\cal C}(L_z) - 1/\tau}\right) \bigg]
\end{split}\ee
with ${\cal C}$ defined in \req{lcdef} and
\be\label{bdef}
b = -i\w \frac{F_0^\sigma}{1+F_0^\sigma} + \frac{1}{\tau}
\ee
Performing the integration and the summation I arrive at:
\be\begin{split}
\Big(&b^\sigma(\w;H) - b^g(\w;H) \Big) - \Big(b^\sigma(\w;0) - b^g(\w;0)\Big) = \\
&-\frac{1}{8\pi^2 D}\bigg[\frac{1}{1+F^\sigma_0} \ln \left| 1- \frac{E_Z^2}{\w^2}\right|
-\ln \left| 1-\frac{E_Z^{*2}}{\w^2}\right| \\
& + \tau \pi \big(|\w| - E_Z^* \big) \Big(\theta(E_Z^2-\w^2)- \theta(E_Z^{*2}-\w^2) \Big) \phantom{\bigg|} \\
& - \left(\frac{F_0^\sigma}{1+F_0^\sigma}\right)^2 \pi \tau |\w| \theta(E_Z^2 - \w^2)
\bigg] \, .
\end{split}\ee
Next, I substitute this result into \req{enden} and then into \req{sphe}; taking the temperature
derivative and rescaling the frequency [$\w \to 2\pi T\w$] I finally obtain:
\be\label{dcvres}\begin{split}
&\delta\cV(H) - \delta\cV(0) = \\
&\frac{1}{4}\frac{T}{D} \left[ I_1\left(\frac{E_Z^*}{2\pi T}\right) -
\frac{1}{1+F_0^\sigma} I_1 \left(\frac{E_Z}{2\pi T}\right) \right] \\
& + \frac{1}{\pi}\frac{T}{D} \left(T\tau\right) \bigg[ \left(\frac{F_0^\sigma}{1+F_0^\sigma}\right)^2
f_3\left(\frac{E_Z}{2T}\right)
+ f_3\left(\frac{E_Z^*}{2T}\right)  \\ & - f_3\left(\frac{E_Z}{2T}\right)
 -\left(\frac{E_Z^*}{2T}\right)\! \bigg(f_2\left(\frac{E_Z^*}{2T}\right)
- f_2\left(\frac{E_Z}{2T}\right)\!\bigg)\bigg]
\end{split}\ee
with $I_1$ defined in \req{I1def} and
\be\label{fidef}
f_n(z) = \int_0^z\!\!d\w \, \frac{\w^n}{\sinh^2 \w} \, .
\ee
The functions $f_n$ can be given in terms of polylogarithms as:
\begin{eqnarray}
f_2(z) &=& \frac{\pi^2}{6} + 2z \log \left( 1-e^{-2z}\right) - \mathrm{Li}_2 \left(e^{-2z}\right) \nonumber \\
&& + z^2 \left(1 - \coth z \right) \nonumber \\
f_3(z) &=& \frac{3}{2} \zeta (3) + 3z^2 \log \left(1-e^{-2z}\right) -3z \mathrm{Li}_2 \left(e^{-2z}\right) \nonumber \\
&& -\frac{3}{2} \mathrm{Li}_3 \left(e^{-2z}\right) + z^3 \left(1 - \coth z \right)
\end{eqnarray}

The second line in \req{dcvres} is the correction to the specific heat in the diffusive limit, while
the last two lines become dominant in the quasiballistic limit. In this limit and
for weak interaction $|F_0^\sigma |\ll 1$, \req{dcvres} reproduces the result of \ocite{chubukov}.

\section{Derivation of the kernel~$\Delta{\cal B}^1$.}
\label{b1der}

In this appendix I briefly outline how to derive the kernel
$\Delta{\cal B}^1$ given in \req{b1eq} starting from the results of Sec.~6.2 of
\ocite{CA}. There, the exact (at linear order in $\nabla T$) solution of the kinetic
equation was given; this solution is unaffected by the parallel field. In the
quasiballistic regime, the main corrections to the thermal conductivity were found to
originate from the term in the bosonic distribution function defined as $\delta N^1$ and which can
be neglected in the diffusive limit;
these corrections are given in Eqs.~(6.35c) and (6.36c) of \ocite{CA}. Performing
the angular integrations in those equations results in Eqs.~(6.38a) and (6.38b) respectively;
by repeating those calculations using the propagators in \reqs{ls}-\rref{lg} a similar result
is obtained in the presence of the parallel field. The explicit
expression is simply found by redefining some of the
quantities appearing on the right hand sides of Eqs.(6.38) as follows: for the
function ${\cal C}$, I substitute the function ${\cal C}(L_z)$ given in \req{lcdef};
the parameter $b'$ is now:
\[
b'(L_z) = -i\big(\w - L_zE_Z^*\big) +\frac{1}{\tau} \, ;
\]
all the other quantities, namely $b$ [\req{bdef}] and
\[
\tilde{N} = v_F \tau \frac{\w^2}{T}\frac{\partial N_P}{\partial \w} \frac{F_0^\sigma}{1+F_0^\sigma}
\]
are unchanged.

{\setlength\arraycolsep{0pt}
To arrive at the kernel $\Delta{\cal B}^1$, the sum over $L_z$ need to be performed,
along with the remaining integral over the magnitude of the momentum $\q$. This integral was
performed in \ocite{CA} with logarithmic accuracy; in the present case however both the infrared
divergence (due to the long-range part of the Coulomb interaction) and the ultraviolet divergence
are absent -- the latter because I consider here the kernel difference, the former because
the triplet channel interaction is short-range -- and I proceed in a different manner.
By rescaling all the dimensionful quantities $\w$, $E_Z^*$, $1/\tau$ and $v_F \q$ by the
temperature $T$, I find that the contributions due to Eq.~(6.38a) are smaller than those
of Eq.~(6.38b) by $1/T\tau$ and can therefore be neglected.
Similarly, to find the leading term in the (Laurent)
expansion over $1/T\tau$ one can set $1/\tau \to 0$ in Eq.~(6.38b). Following this procedure,
I get:\begin{eqnarray*}
&&\int\!\frac{d^2q}{(2\pi)^2} \!\int\!\frac{d\nn_1 d\nn_2}{\W_2} n_{1\mu}
\Re \left\{{\cal L}^\sigma(L_z) \delta_\nu N^1_{12} \right\} \\
&& \simeq -\delta_{\mu\nu}\frac{\tau\w}{8\pi v_F^2} \frac{F_0^\sigma}{1+F_0^\sigma} \tilde{N}\!
\int_0^\infty\!\!\!dp\Bigg[ \Re \frac{1}{\sqrt{p-(1-L_zE_Z^*/\w)^2}} \\
&& \qquad \ \times \Re \frac{1}{\sqrt{p - (1-L_zE_Z^*/\w)^2} - i\frac{F_0^\sigma}{1+F_0^\sigma}}
\\ && \qquad \ \times
\left(1-\frac{(1-L_zE_Z^*/\w)^2}{p}\right) \Bigg] \, ,
\end{eqnarray*}
where in terms of the original variables $p=(v_Fq/\w)^2$, and the approximate equality
indicates that I am neglecting higher order terms in $1/T\tau$.
The integral over $p$ is logarithmically
divergent, but the difference between the above expression and the similar one at zero parallel field
is finite. The exact integration of this difference gives finally the expression for the kernel
$\Delta{\cal B}^1$ given in \reqs{b1eqs}.}

\section{Parallel field dependence of the electrical conductivity.}
\label{app}

The aim of this appendix is to compare the present approach to the one of \ocite{znapar} for
the calculation of the parallel field magneto-conductivity $\Delta\sigma$, which is given by:
\be\label{mc}
\Delta\sigma = \sd \int\!d\w\Big[\Delta{\cal E}(\w) + \Delta{\cal S}^{el}(\w)\Big]
\frac{\partial}{\partial\w} \Big[\w N_P(\w) \Big] \, .
\ee
This formula follows from Eq.~(6.8) of \ocite{CA}, and the kernels are defined in the subsequent
Eq.~(6.9).

In the diffusive limit only $\Delta{\cal E}$ is relevant, since the
kernel ${\cal S}^{el}$ gives contributions smaller by the factor $T\tau \ll 1$, as discussed
in \ocite{CA}; it is straightforward to verify that substituting
\req{Deker} into \req{mc}, the diffusive limit result of \ocite{znapar} is recovered.
Viceversa, it can be shown that in the quasiballistic limit the kernel ${\cal E}$ can be neglected
since larger corrections are due to ${\cal S}^{el}$ -- this can be done for example
by rescaling all dimensionful quantities by the temperature, as described in Appendix~\ref{b1der}.
Proceeding as detailed there, i.e. dropping higher order terms in $1/T\tau$, and introducing
the shorthand notation:
\[
\s(a|b) \equiv \sgn(a) - \sgn(b) \, ,
\]
I obtain for the kernel $\Delta{\cal S}^{el}$ in the quasiballistic regime:
\begin{subequations}\label{Dsker}
\be
\Delta{\cal S}^{el} = \Delta{\cal S}^{11} + \Delta{\cal S}^{12}
\ee
\begin{eqnarray}
\Delta{\cal S}^{11} &\simeq & \frac{e^2}{\sd 2\pi^2} \frac{\tau}{\w} \frac{\pi}{2} \sum_{L_{z}=\pm1}
\bigg[  \s (\w-L_zE_Z|\w-L_zE_Z^*) \nonumber \\ && \times \left(\w - L_zE_Z^*\right) -
\w \frac{F_0^\sigma}{1+F_0^\sigma} \s(\w-L_zE_Z|\w)\bigg] \nonumber \\
\end{eqnarray}
\begin{eqnarray}
\Delta{\cal S}^{12} &\simeq & \frac{e^2}{\sd 2\pi^2} \frac{\tau}{\w} \frac{\pi}{2} \sum_{L_{z}=\pm1}
\bigg[  \s (\w-L_zE_Z|\w-L_zE_Z^*) \nonumber \\
&& \times \left(\w - L_zE_Z^*\right)\frac{2\w F_0^\sigma}{\w(1+2F_0^\sigma) - L_zE_Z} \nonumber
\\ && - \w \frac{F_0^\sigma}{1+F_0^\sigma} \s(\w-E_Z|\w)\bigg] \, .
\end{eqnarray}\end{subequations}

\reqs{mc} and \rref{Dsker} lead to the quasiballistic limit result of \ocite{znapar}; in particular,
the sum of the first terms in square brackets gives the contribution denoted there by $K_2$, while the
remaining terms give the $K_1$ contribution. Here I point out that in the low-field limit
$E_Z, \, E_Z^*\ll 2T$, the quasiballistic magneto-conductance is given by:
\be
\Delta\sigma \approx \frac{e^2}{\pi}\frac{2F_0^\sigma}{1+F_0^\sigma} T\tau
\frac{1}{3}\left(\frac{E_Z}{2T}\right)^2 f(F_0^\sigma) \, ,
\ee
where
\begin{eqnarray}\label{fcorr}
f(z) = 1-\frac{z}{1+z} \bigg[ 1 &+& \frac{1}{2} \frac{1}{1+2z} + \frac{1}{(1+2z)^2} \\ &-&
\frac{2(1+z)\ln 2(1+z)}{(1+2z)^3} \bigg] \, . \nonumber
\end{eqnarray}
This expression corrects the wrong definition of $f(z)$ given after Eq.~(14) of
\ocite{znapar}.\footnote{The erroneous definition in \ocite{znapar} is probably the consequence
of an algebraic mistake. As a further check, I have verified numerically that \req{fcorr} is indeed
the right formula.} While the difference between the two definitions is numerically small
(less than 4\%) for $-0.57\lesssim F_0^\sigma \lesssim 0.14$, it grows rapidly outside this parameter
range, and use of \req{fcorr} rather than its counterpart of \ocite{znapar} may be important in
a comparison between theory and experiment.

\end{document}